\documentclass[12pt]{article}
\usepackage{amssymb, amsmath, bm, float, epsfig, color}
\usepackage[centering]{geometry}
\makeatletter
    
    \@addtoreset{equation}{section}
 \makeatother

\definecolor{mygrn}{rgb}{0,0.5,0}

\begin{document}
\title{
\vspace{-30pt}
\begin{flushright}
\normalsize WU-HEP-18-06 \\
DESY 18-092 \\*[55pt]
\end{flushright}
{\Large \bf Effects of localized $\mu$-terms at the 
fixed points in magnetized orbifold models\\*[20pt]}
}

\author{
Hiroyuki~Abe,$^1$\footnote{E-mail address: abe@waseda.jp} \quad
Makoto~Ishida,$^1$\footnote{E-mail address: ncanis3@fuji.waseda.jp} \, and \, 
Yoshiyuki~Tatsuta$^{2}$\footnote{E-mail address: yoshiyuki.tatsuta@desy.de}\\*[30pt]
$^1${\it \normalsize Department of Physics, Waseda University, Tokyo 169-8555, Japan}\\
$^2${\it \normalsize Deutsches Elektronen-Synchrotron DESY, Hamburg 22607, Germany}\\*[55pt]
}

\date{
\centerline{\small \bf Abstract}
\begin{minipage}{0.9\textwidth}
\medskip\medskip 
\small
We consider magnetized orbifolds, where the supersymmetric mass term for a pair of up- and down-type Higgs (super)fields, called $\mu$-term, is localized at the orbifold fixed points, and study the effects on the zero-mode spectra. The zero-mode degeneracy to be identified as the generation in four-dimensional (4D) effective theories is determined by the magnetic fluxes. It is known that multiple Higgs zero-modes appear in general in magnetized orbifold models. We derive the analytic form of the $\mu$-term matrix in the 4D effective theory generated by the localized sources on $T^2/Z_2$ orbifold fixed points, and find that this matrix can lead to a distinctive pattern of the eigenvalues that yields hierarchical $\mu$-terms for the multiple Higgs fields. The lightest ones can be exponentially suppressed due to the localized wavefunctions of zero-modes determined by the fluxes, while the others are of the order of the compactification scale, which can provide a dynamical origin of the electroweak scale as well as a simultaneous decoupling of extra Higgs fields. We also show that a certain linear combination of the lightest Higgs fields could generate the observed mass ratios of down-type quarks through their Yukawa couplings determined by the wavefunctions.
\end{minipage}}

\begin{titlepage}
\maketitle
\thispagestyle{empty}
\end{titlepage}
\tableofcontents

\section{Introduction}
In particle physics models constructed so far, extra-dimensional theories sometimes play important roles for explaining the observed structures of the standard model (SM). In particular, revealing the origins of chiral structure, generation structure, and flavor structure such as the observed hierarchical masses and mixing angles of quarks and leptons are challenging issues in theories beyond the SM. The introduction of magnetic fluxes in extra-dimensional space is a promising way to realize the above SM structures in four-dimensional (4D) effective theories (see {Refs.~\cite{Bachas:1995ik,Cremades:2004wa,Ibanez:2012zz}} and references therein). In such theories, the number of degenerated chiral zero-modes, those arise as a consequence of magnetic fluxes and can be recognized as generations, is determined by the number of fluxes they feel. 

The orbifold projection is another way to obtain chiral structures. When we consider orbifold models with magnetic fluxes, the relation between the number of magnetic fluxes and the number of chiral zero-modes changes drastically~\cite{Braun:2006se,Abe:2008fi}, providing wide varieties of phenomenological models. A series of studies about the phenomenological model building on magnetized orbifolds was carried out in Refs.~\cite{Abe:2008fi,Abe:2008sx,Abe:2009uz,Abe:2012ya,Abe:2012fj,Abe:2013bba,Abe:2014nla,Abe:2014soa,Abe:2014vza,Buchmuller:2015eya,Buchmuller:2015jna,Abe:2016jsb,Kobayashi:2016qag}, the generalizations and classifications of such models have been done in Refs.~\cite{Fujimoto:2013xha,Abe:2013bca,Abe:2014noa,Abe:2015yva,Fujimoto:2016zjs} {and the investigations on quantum corrections \cite{Buchmuller:2016gib,Buchmuller:2018eog,Ghilencea:2017jmh}}. In magnetized orbifold models, multiple Higgs generations also appear in general. Especially, due to the gauge invariance, the Higgs multiplicity is unavoidable if we require the nonvanishing Yukawa couplings to three generations of quarks and leptons, those arise as degenerated zero-modes. Therefore, it is nontrivial to decouple extra Higgs generations, in order the multiple Higgs models to be phenomenologically viable. 

The supersymmetry is also introduced in many models {beyond} the SM (see Ref.~\cite{Martin:1997ns} for a review). It protects scalar masses against huge radiative corrections, hence gives a solution to the fine-tuning problem of Higgs mass parameter in the SM. The minimal supersymmetric standard model (MSSM) predicts a unification of the SM gauge couplings at a high energy scale $M_U \sim 10^{16}$ GeV, which may indicate a grand unification of the SM gauge groups. The Higgs sector of the MSSM consists of two Higgs doublets~$h_u$ and $h_d$, those form chiral multiplets under the supersymmetry and described by chiral superfields $H_u$ and $H_d$, respectively. The Higgs potential responsible for the electroweak symmetry breaking is characterized by a supersymmetric mass parameter~$\mu$ appears in the so-called $\mu$-term written in the superspace as
\begin{eqnarray}
\mathcal{L}_{\rm Higgs} 
&=& \int d^2\theta\left[\mu H_uH_d+{\rm h.c.}\right], 
\label{mu-term}
\end{eqnarray}
as well as the gauge couplings and soft supersymmetry breaking parameters. It is known that the mass parameter~$\mu$ should be of ${\cal O}(10^2)$ GeV to reproduce the correct electroweak scale required for explaining the observed masses of electroweak gauge bosons, otherwise a fine-tuning is required between the supersymmetric mass parameter~$\mu$ and certain supersymmetry breaking mass parameters. However, the reason why $\mu$ is so small compared with the fundamental scale such as the Planck scale $M_P \sim 10^{18}$ GeV or the unification scale $M_U \sim 10^{16}$ GeV has never been understood well. This is referred to as the $\mu$-problem (see Ref.~\cite{Martin:1997ns} for a review). 

The supersymmetry in a higher-dimensional spacetime is larger than the one in four-dimensions. It sometimes occurs that the $\mu$-term~(\ref{mu-term}) is forbidden by such an extended supersymmetry. Then, it is problematic when we construct supersymmetric models in higher-dimensional spacetime. In orbifold models, however, there are fixed points in extra-dimensional space and symmetries such as gauge symmetries and/or supersymmetry are partially broken at these points. Hence there is a possibility that some Lagrangian terms, those are forbidden in the bulk, are allowed at the fixed points where the symmetries are reduced. For example, certain mass terms localized at orbifold fixed points {coexisting} with bulk magnetic fluxes were studied in Ref.~\cite{Ishida:2017avx} and applied to a neutrino phenomenology in Ref.~\cite{Ishida:2018bbl}. Then, it is certainly interesting to consider localized $\mu$-terms at the fixed points of magnetized orbifolds. 

In this paper, we study the effects of $\mu$-terms localized at the fixed points on Kaluza-Klein (KK) zero-modes of Higgs (super)fields in magnetized orbifold models. We consider a six-dimensional (6D) spacetime\footnote{We assume that the 6D models considered in this paper can be embedded into higher dimensions to preserve the supersymmetry even with magnetic fluxes, that is explained in some detail in the next section.}, for simplicity, where two extra dimensions are compactified on a toroidal orbifold~$T^2/Z_2$ which has four fixed points. By introducing magnetic fluxes and the localized $\mu$-terms, we derive the {effective} $\mu$-terms for the degenerated KK zero-modes of Higgs (super)fields, and analyze their mass matrix deformed by the localized sources. We will find that the rank of the mass matrix is determined by the number of fixed points where nonvanishing $\mu$-terms are located, and the eigenvalues and the eigenstates have phenomenologically interesting features, depending on the number of magnetic fluxes felt by the Higgs fields, their orbifold parities and the complex structure of the torus~$T^2$. 

The organization of this paper is as follows. In Sec.~\ref{review}, we review magnetized orbifold models on $T^2/Z_2$. In Sec.~\ref{class1}, the effects of localized $\mu$-terms on Higgs zero-modes are studied in the case that the number of KK zero-modes is less than or equal to the number of {localized} sources. We will find that one of the mass eigenvalues tends to be hierarchically small compared to the others in this case. In Sec.~\ref{class2}, the opposite case to the one analyzed in Sec.~\ref{class1} is studied, and a model is shown to realize the observed mass hierarchy for down-type quarks. In Sec.~\ref{sec:KKeffects}, we confirm that the effects from KK excitation modes are negligible in the zero-mode analyses. Sec.~\ref{conclusion} is devoted to conclusions.

\section{Review of magnetized orbifold models}
\label{review}
In this section we review extra-dimensional models with magnetic fluxes on a torus or a toroidal orbifold based on Refs.~\cite{Cremades:2004wa,Abe:2008fi}. In this paper, we consider 6D spacetime, i.e., the observed 4D Minkowski spacetime~$\mathcal{M}^4$ and the torus~$T^2$ or orbifold~$T^2/Z_2$ as a two-dimensional extra space. Their coordinates are represented by $x^{\mu}$ ($\mu=0,1,2,3$) and $z=(y^5+\tau y^6)/2\pi R$, respectively, where $2\pi R$ is the period of $T^2$ in $y^5$-direction, and 
$\tau\in\mathbb{C} \,\, {({\rm Im} \, \tau > 0)}$ parameterizes the complex structure of $T^2$.
\footnote{{Indeed, the complex structure $\tau$ is a scalar field (one of the moduli fields) and its VEV should be dynamically stabilized in the context of supergravity.
However, we assume an appropriate moduli stabilization and consider the complex structure $\tau$ as a parameter in this paper.}}
We call $M_c=1/R$ the compactification scale in this paper. 

First, we consider the torus~$T^2$ defined by the identification 
\begin{eqnarray}
z &\sim& z+1 \ \sim \ z+\tau, 
\label{T2periods}
\end{eqnarray}
and introduce background magnetic fluxes~$b$ for an Abelian gauge field strength~$F$ such as
\begin{eqnarray}
b &=& \int_{T^2}F,\qquad F \ = \ \frac{ib}{2{\rm Im}\,\tau}dz\wedge d\bar{z}.
\end{eqnarray}
Hence, the extra-dimensional components~$A_z$ of the vector potential can be written as
\begin{eqnarray}
A_z &=& \frac{ib}{2{\rm Im}\,\tau}{\rm Im}\left(\bar{z}dz\right).
\end{eqnarray}
For the single-valuedness of charged-matter wavefunction up to a gauge transformation, the value of $M=Qb/2\pi$ is restricted to be an integer, that is, Dirac's quantization condition, where $Q$ is the $U(1)$ charge of the corresponding matter field. 
Under an appropriate twisted boundary condition determined by the flux~$b$, we can expand 6D charged spinor and scalar fields $\Psi(x^\mu,z)$ and $\Phi(x^\mu,z)$ into Kaluza-Klein (KK) modes, respectively, as 
\begin{eqnarray}
\label{fermionKK}
\Psi\left(x^\mu,z\right) 
&=& \sum_{n,{J}}\psi_n^J\left(x^\mu\right)\otimes\chi_n^J(z), \\
\label{scalarKK}
\Phi\left(x^\mu,z\right) 
&=& \sum_{n,{J}}\phi_n^J\left(x^\mu\right)\otimes\varphi_n^J(z),
\end{eqnarray}
where $n$ represents KK modes including massless zero-modes ($n=0$) and an infinite tower of massive higher-modes ($n>0$). The additional index $J$ expresses the degeneration of each KK mode explained below.  

We describe the zero-mode wavefunctions~$\chi_0^J(z)$ for the spinor field~$\Psi\left(x^\mu,z\right)$ in Eq.(\ref{fermionKK}) as eigenstates of the Dirac operator on $T^2$ defined with the covariant derivative $D_z=\partial_z-iQA_z$, those are {denoted} in the chirality basis by $\chi_0^J=\left(\psi_+^J,\psi_-^J\right)^T$. For $M=Qb/2\pi>0$, they can be written analytically as $\psi_+^J(z)=\Theta^{J,M}_{T^2}(z)$, where 
\begin{eqnarray}
\Theta^{J,M}_{T^2}(z) \ \equiv \ {\mathcal{N}_M} \, 
e^{i\pi Mz {\rm Im}\, z /{\rm Im}\,\tau} \cdot \vartheta
\begin{bmatrix}
J/M\\[5pt]
0 
\end{bmatrix}
\left(Mz,M\tau\right), 
\label{wavefuncT2}
\end{eqnarray}
and $J=0, 1,2, \cdots, M-1$ labels $M$ degenerated zero-modes.
In this expression, $\vartheta$ describes the Jacobi-theta function and 
${{\cal N}_M} = (2{{\rm Im}\,\tau}M/{\cal A}^2)^{1/4}$ 
is a normalization constant (with the mass dimension~$+1$) expressed by the area of torus~$\mathcal{A}$. 
We should comment that there is no normalizable solution for $\psi_-$ for $M>0$. This means that 4D effective theories possess a chiral structure by introducing a non-zero magnetic flux. In addition to that, the chiral zero-modes~$\psi_+$ have the $M$-degeneration labeled by $J$. (For $M<0$, there is no normalizable solution for $\psi_+$, and $|M|$-degenerated chiral zero-modes~$\psi_-$ appear.) 
We can interpret it as a generation structure of the chiral fermion in 4D effective theories, which can be the origin of, e.g., three families of quarks and leptons in the standard model with a suitable choice of fluxes. 


{
In a similar way, we can analyze KK modes for the scalar field~$\Phi\left(x^\mu,z\right)$ in Eq.~(\ref{scalarKK}) as eigenstates of the Laplace operator on $T^2$. 
{By introducing non-zero magnetic fluxes on $T^2$ or $T^2/Z_2$, 6D scalar fields have no massless mode contrary to the spinor fields and the original 6D ${\cal N}=1$ supersymmetry is completely broken}~\cite{Cremades:2004wa,Hamada:2012wj}. 
On the other hand, in this paper, we consider the situation that the orbifold~$T^2/Z_2$ of our 6D models are somehow embedded into a higher-dimensional space, and our 6D scalars are outer extra-dimensional components of higher-dimensional vectors. In this case, fluxes and/or curvatures outside the $T^2/Z_2$ can provide both positive and negative contributions to the scalar mass squared (called twist in Ref.~\cite{Conlon:2008qi}). We assume that these contributions cancel the scalar masses on $T^2/Z_2$ and the lowest modes totally become massless preserving the 4D ${\cal N}=1$ supersymmetry.\footnote{A typical example for such a higher-dimensional completion is the ten-dimensional (10D) super Yang-Mills (SYM) theory compactified on $(T^2)^3/(Z_2 \times Z'_2)$, where the fluxes on three tori cancel with each other in the ${\cal N}=1$ supersymmetry transformations~\cite{Cremades:2004wa}.} Then, the scalar mode analysis is completely analogous to the spinor one and both of them can be simultaneously performed on ${\cal N}=1$ superspace~\cite{Abe:2012ya,Abe:2015jqa}. We remark that the fixed points on $T^2/Z_2$ in our 6D models should be considered as those in the total higher-dimensional space at which the supersymmetry is reduced to 4D ${\cal N}=1$, allowing the $\mu$-terms we introduce later. Even in such a higher-dimensional space, we can focus on its $T^2/Z_2$ subspace for our purpose if a single zero-mode (no degeneracy) for each Higgs (super)fields arises on the extra space outside $T^2/Z_2$ and its wavefunction is not severely localized in the outer space. In this case, we expect that just overall factors of the terms in the 4D effective action will be slightly affected by the extra space outside $T^2/Z_2$.\footnote{A typical example for such a case is again the 10D SYM compactified on $(T^2)^3/(Z_2 \times Z'_2)$ mentioned in the previous footnote. In the phenomenological model building, the generation structure should arise from one of the three tori in order the Yukawa matrices to become full rank for three generations of quarks and leptons, that may be identified as the $T^2/Z_2$ focused on in this paper.} With these assumptions, we will perform model-independent analyses on the $T^2/Z_2$ subspace in the next section without specifying the outer extra space.

Given the wavefunctions for spinor and scalar zero-modes on $T^2$ in the form of Eq.~(\ref{wavefuncT2}), Yukawa coupling constants in the 4D effective theory can be expressed analytically as~\cite{Cremades:2004wa} 
\begin{eqnarray}
Y^{IJK}&=&g\int_{T^2}d^2z\,
\Theta^{I,M_1}_{T^2}(z) 
\Theta^{J,M_2}_{T^2}(z) 
\left(\Theta^{K,M_3}_{T^2}(z)\right)^*\nonumber\\
&=& g\,
{\frac{ \mathcal{N}_{|M_1|} \mathcal{N}_{|M_2|} }{ \mathcal{N}_{|M_3|}}} \, 
\sum_{m=0}^{|M_3|-1}\vartheta
\begin{bmatrix}
\frac{M_2I -M_1J+M_1M_2m}{M_1M_2M_3}\\[5pt]
0 
\end{bmatrix}
(0,\tau M_1M_2M_3)\times\delta_{I+J+M_1m,K+M_3l},\ 
\label{T2yukawa}
\end{eqnarray}
where the gauge invariance requires $M_1+M_2=M_3$ and the overall constant~$g$ has the  mass dimension $-1$. Note that the Yukawa coupling originates from a gauge coupling if 
the 6D models are embedded in 10D super Yang-Mills (SYM) theory. 
 
Now, we extend the torus~$T^2$ to the orbifold~$T^2/Z_2$. In this case, a further identification on the extra coordinate, called $Z_2$-projection, 
\begin{eqnarray}
z &\sim& -z, 
\label{z2projection}
\end{eqnarray}
is imposed in addition to the toroidal ones~(\ref{T2periods}), and the resultant fundamental region on $z$-plane is shown in {Fig.~\ref{orbifoldregion}}. 
\begin{figure}[t]
\centering
\includegraphics[width=0.75\textwidth]{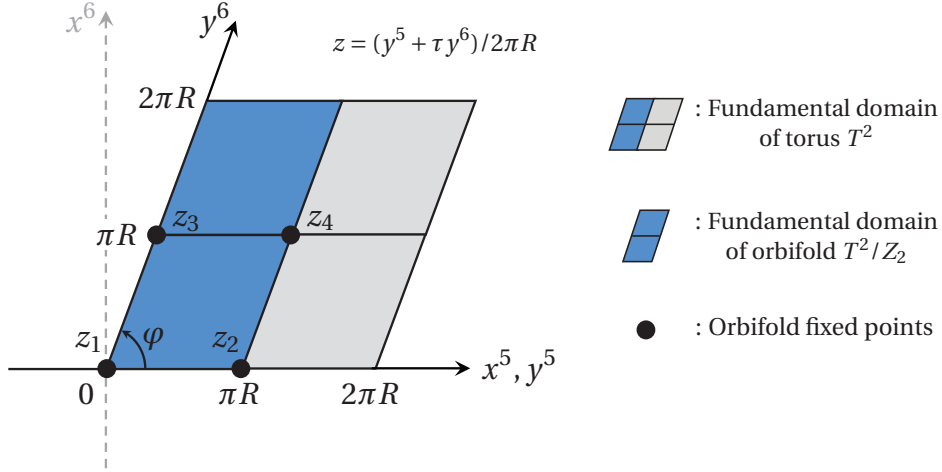}
\caption{The fundamental domain of the torus $T^2$ and the orbifold $T^2/Z_2$. 
The {blue/gray} region is a fundamental region of $T^2$, the {blue} region is a fundamental region of $T^2/Z_2$, and the {black} dots represent the four fixed points of the orbifold.
This figure is drawn for a generic complex structure $\tau \in \mathbb{C}$, where {the angle~$\varphi$} between $y^5$ and $y^6$ axes is given by {$\tau\equiv|\tau|e^{i\varphi}$}.}
\label{orbifoldregion}
\end{figure}
Due to this projection, wavefunctions on $T^2/Z_2$ can be expressed as certain linear combinations of the wavefunctions~(\ref{wavefuncT2}) on $T^2$ shown previously. Those are described by 
\begin{eqnarray}
\Theta^{J,M}_{T^2/Z_2}(z) 
&=& \frac{1}{\sqrt{2}}\left[\Theta^{J,M}_{T^2}(z)+\eta\Theta^{M-{J},M}_{T^2}(z)\right], 
\label{wavefuncT2Z2}
\end{eqnarray}
satisfying the boundary condition 
$\Theta^{J,M}_{T^2/Z_2}\left(-z\right)=\eta\Theta^{J,M}_{T^2/Z_2}\left(z\right)$, 
where $\eta=\pm 1$ is called the $Z_2$-parity. The wavefunctions on $T^2/Z_2$ are classified into either the parity even mode with $\eta=+1$ or the odd mode with $\eta=-1$, under the $Z_2$-reflection. The relation between the number~$M$ of fluxes on $T^2/Z_2$ and the degeneracy~$M_\eta$ of parity even~($\eta=+1$) or odd~($\eta=-1$) zero-modes is shown in Tab.~\ref{T2Z2generation}. 
\begin{table}[t]
\begin{center}
\begin{tabular}{|c|cccccccccc|cc|}\hline
$M$ & $0$ & $1$ & $2$ & $3$ & $4$ & $5$ & $6$ & $7$ & $8$ & $\cdots$ & $2{l}$ & $2{l}+1$\\ \hline
$M_\eta=M_+$ & $1$ & $1$ & $2$ & $2$ & $3$ & $3$ & $4$ & $4$ & $5$ & $\cdots$ & ${l}+1$ & ${l}+1$\\
$M_\eta=M_-$ & $0$ & $0$ & $0$ & $1$ & $1$ & $2$ & $2$ & $3$ & $3$ & $\cdots$ & ${l}-1$ & ${l}$ \\ \hline
\end{tabular}
\caption{The relation between the number~$M$ of fluxes felt by a field on $T^2/Z_2$ and the degeneracy~$M_\eta$ of parity even~($\eta=+1$) or odd~($\eta=-1$) zero-modes of the field, where ${l}=0,1,2,\cdots$. We find $M=M_++M_-$.}
\label{T2Z2generation}
\end{center}
\end{table}
According to this structure, contrary to the toroidal case, the number~$M_\eta$ of zero-modes with each parity~$\eta=\pm 1$ can be different from the number~$M$ of fluxes, $M_\eta \ne M$ in general, depending on their parity. Therefore, we can construct a wide variety of models on orbifolds. (See Refs.~\cite{Braun:2006se,Abe:2008fi,Abe:2008sx,Abe:2013bca} and subsequent papers.)

In the next two sections, we will analyze $\mu$-terms~(\ref{mu-term}) localized on the orbifold fixed points. Hence, we discuss properties of the wavefunction~(\ref{wavefuncT2}) on the $T^2/Z_2$ fixed points here. As shown in {Fig~\ref{orbifoldregion}}, there are four fixed points $k=1,2,3,4$ located at $z=z_k$ on the $T^2/Z_2$ orbifold, where $z_k$ are {given} by 
\begin{eqnarray}
\left(z_1,z_2,z_3,z_4\right) 
&=& \left(0,\frac{1}{2},\frac{\tau}{2},\frac{1+\tau}{2}\right). 
\label{FP}
\end{eqnarray}
In the following analyses, {we parameterize them by}
\begin{eqnarray}
z_k &=& \frac{p+q\tau}{2}, 
\label{pqdef}
\end{eqnarray}
where $p,q=0,1$, hence four combinations $(p,q)=(0,0)$, $(0,1)$, $(1,0)$ and $(1,1)$ correspond to $z_1$, $z_2$, $z_3$ and $z_4$, respectively. 

According to Ref.~\cite{Ishida:2017avx}, the values of wavefunctions~(\ref{wavefuncT2}) on $T^2$ at these fixed points $k=1,2,3,4$ are computed as 
\begin{eqnarray}
\Theta^{M-{J},M}_{T^2}(z_k) &=& (-1)^{M\delta_{k,4}}\Theta^{J,M}_{T^2}(z_k).
\end{eqnarray}
Then, we find the expression for wavefunctions~(\ref{wavefuncT2Z2}) on $T^2/Z_2$ as 
\begin{eqnarray}
\Theta^{J,M}_{T^2/Z_2}(z_k)
&=& \frac{1}{\sqrt{2}}\left[1+\eta(-1)^{M\delta_{k,4}}\right]\Theta^{J,M}_{T^2}(z_k) 
\ \equiv \ \rho_k^M\Theta^{J,M}_{T^2}(z_k),
\label{nazo}
\end{eqnarray}
where 
\begin{eqnarray}
\rho_k^M=\left\{
\begin{array}{cc}
0\ &\ (M:{\rm odd}\ {\rm and}\ k=4)\\
\sqrt{2}\ &\ ({\rm the\ other\ cases}) 
\end{array}
\right.
\label{rho}
\end{eqnarray}
{for $\eta=+1$.} This tells us that, for given $M$ and $J$, the value of orbifold wavefunction at any fixed point is proportional to the one of torus {wavefunctions} at the same fixed point. Note that when a wavefunction has the even parity ($\eta=+1$), the coefficient $\rho_k^M$ vanishes only if $M$ is an odd integer and $k$ is four. Therefore, if we introduce any odd number of fluxes, the parity even wavefunctions vanish at the fixed point~$z_4$, while the parity-odd wavefunctions do not. We must be careful {of} this property when we construct models.

\section{Deformation of Higgs zero-modes by localized $\mu$-terms}
\label{class1}
The supersymmetry in higher-dimensional spacetime corresponds to ${\cal N} \ge 2$ in terms of the 4D supercharge. It sometimes occurs in the model building that the terms written in ${\cal N}=1$ superpotential, like the $\mu$-term, are forbidden in the higher-dimensional bulk action. For example, as mentioned in the previous section, we implicitly assume that our 6D models are embedded in 10D spacetime, where the (global) supersymmetry is enhanced to ${\cal N}=4$, and the $\mu$-term is actually forbidden in the bulk action of magnetized orbifold models based on 10D SYM theory. Even in this case, there is a possibility that superpotential terms are allowed at some orbifold fixed points where the supersymmetry is broken down to ${\cal N}=1$. We consider such a situation in this paper, and study the effects of localized $\mu$-terms at the fixed points of a typical orbifold background~$T^2/Z_2$ which could be contained in higher-dimensional orbifolds or (certain singular limits of) some manifolds like Calabi-Yau spaces, in a bottom-up manner. 
In the following, we assume {a K\"ahler potential is} the canonical one in the bulk discussed in Ref.~\cite{ArkaniHamed:2001tb}.

Motivated by the above arguments, we introduce localized $\mu$-terms at the fixed points located at $z=z_k$ ($k=1,2,3,4$) defined in Eq.~(\ref{FP}) on the orbifold $T^2/Z_2$, and study their effects on Higgs zero-modes in the 4D effective theory. 
Such $\mu$-terms are described by the 6D Lagrangian written in 4D ${\cal N}=1$ superspace, 
\begin{eqnarray}
\mathcal{L}_{\mu{\textrm -}{\rm term}} &=& 
\int d^2\theta\sum_{k=1}^4\nu_kH_u(x^\mu,z,\theta)H_d(x^\mu,z,\theta)\delta^{(2)}(z-z_k)+{\rm h.c.}, 
\label{6Dmuterm}
\end{eqnarray}
where $H_u$ and $H_d$ are up- and down-type Higgs chiral superfields {with the mass dimensions $+2$}, $\theta$ is a Grassmann coordinate of the superspace, and $\nu_k\in\mathbb{R}$ are constant parameters with the mass dimension $-1$. 
The gauge invariance of  $\mathcal{L}_{\mu{\textrm -}{\rm term}}$ requires that the numbers of fluxes felt by $H_u$ and $H_d$ have the same magnitude but the opposite sign to each other, which we assign $M>0$ and $-M<0$, respectively. On top of that, if $H_u$ and $H_d$ have different orbifold parities, their zero-mode degeneracies differ from each other as shown in Tab.~\ref{T2Z2generation}, that may cause problems in the low-energy effective theory, e.g., the SM gauge groups become anomalous.\footnote{It is conceivable that such an anomaly would be canceled by Green-Schwarz mechanism~\cite{Green:1984sg}. In this case, bulk wavefunctions of charged matter fields could be deformed depending on the details of anomaly cancellation, which is beyond the scope of this paper.} Therefore, we restrict ourselves to the case that both $H_u$ and $H_d$ have the same parity~$\eta=+1$ or $-1$. 

\subsection{4D effective $\mu$-term}
Because we assume that the ${\cal N}=1$ supersymmetry, here described by $\theta$, is preserved in our system, the superfields $H_u(x^\mu,z,\theta)$ and $H_d(x^\mu,z,\theta)$ themselves can be expanded into the KK towers as in Eq.~(\ref{fermionKK})~\cite{Abe:2012ya,Abe:2015jqa}. In this section and the next section, we focus on the KK zero-modes, neglecting all the KK excitation modes whose masses are of ${\cal O}(M_c)$ by assuming $M_c$ is so high {($\sim 10^{16}$ GeV in this paper)} that they decouple from the zero-modes at a lower energy. Later, in Sec.~\ref{sec:KKeffects}, we will confirm that their effects are actually negligible based on a numerical analysis. Then, the 4D effective $\mu$-terms are described as 
\begin{eqnarray}
\mathcal{L}_{\mu{\textrm -}{\rm term}}^{\rm eff} 
&=& \int d^2\theta\sum_{K,L}\left[\sum_{k=1}^4\nu_k\Theta^{K,M}_{T^2/Z_2}(z_k)\left(\Theta^{L,M}_{T^2/Z_2}(z_k)\right)^*\right]H_u^K(x^\mu,\theta)H_d^L(x^\mu,\theta)+{\rm h.c.} \nonumber \\
&\equiv& \int d^2\theta\sum_{K,L}\mu_{\rm eff}^{KL}H_u^K(x^\mu,\theta)H_d^L(x^\mu,\theta)+{\rm h.c.}, 
\end{eqnarray}
where the 4D superfields $H_u^K(x^\mu,\theta)$ and $H_d^L(x^\mu,\theta)$ with $K,L=0,1,\ldots,M_\eta-1$ describe the degenerated KK zero-modes of 6D superfields $H_u(x^\mu,z,\theta)$ and $H_d(x^\mu,z,\theta)$, respectively, and $\mu_{\rm eff}^{KL}$ is the supersymmetric mass matrix for them, which we call the effective $\mu$-matrix.
\footnote{
{Note that, without the magnetic fluxes, the generation structure does not appear and zero-mode wavefunctions are just constants (unless introducing bulk masses those are prohibited now). In this case, the effective $\mu$-parameter in Eq. (3.2) is of the order of the compactification scale $M_c$, thus the mass of the single Higgs boson is of ${\cal O}(M_c)$ without a fine-tuning.
}
}

{
}

Using the relation~(\ref{nazo}), we can describe the $M_\eta \times M_\eta$ matrix $\mu_{\rm eff}^{KL}$ with respect to the wavefunctions~(\ref{wavefuncT2}) on $T^2$ as\footnote{
As mentioned in the previous section, we need to consider a higher-dimensional embedding of our 6D models to assure the 4D ${\cal N}=1$ supersymmetry with the presence of magnetic fluxes. Even in this case, if a single zero-mode (no degeneracy) for each Higgs (super)fields arises on the extra space outside $T^2/Z_2$ and its wavefunction is not severely localized in the outer space, we expect that just an extra overall factor of ${\cal O}(1)$ appears in the expression of $\mu_{\rm eff}^{KL}$ shown in Eq.~(\ref{mumatrix}), which does not affect the arguments in this paper.} 
\begin{eqnarray}
\mu_{\rm eff}^{KL} &=& \sum_{k=1}^4\nu_k\left(\rho_k^M\right)^2\Theta^{K,M}_{T^2}(z_k)\left(\Theta^{L,M}_{T^2}(z_k)\right)^*.
\label{mumatrix}
\end{eqnarray}
{In the following analyses, we consider all possible combinations of introducing non-zero values for $\nu_k$ shown in  Tab. 2. For example, Case 1 implies a situation where only $\nu_1$ takes a non-zero value at the first fixed point of $T^2/Z_2$ and other coefficients of localized $\mu$-terms are vanishing at the other fixed points.}
%
\begin{table}[t]
\begin{center}
\begin{tabular}{|c|ccccccccccccccc|}\hline
Case & $1$ & $2$ & $3$ & $4$ & $5$ & $6$ & $7$ & $8$ & $9$ & $10$ & $11$ & $12$ & $13$ & $14$ & $15$\\ \hline
$\nu_1$ & $\checkmark$ & \ & \ & \ & $\checkmark$ & $\checkmark$ & $\checkmark$ & \ & \ & \ & $\checkmark$ & $\checkmark$ & $\checkmark$ & \ & $\checkmark$ \\
$\nu_2$ & \ & $\checkmark$ & \ & \ & $\checkmark$ & \ & \ & $\checkmark$ & $\checkmark$ & \ & $\checkmark$ & $\checkmark$ & \ & $\checkmark$ & $\checkmark$ \\
$\nu_3$ & \ & \ & $\checkmark$ & \ & \ & $\checkmark$ & \ & $\checkmark$ & \ & $\checkmark$ & $\checkmark$ & \ & $\checkmark$ & $\checkmark$ & $\checkmark$ \\
$\nu_4$ & \ & \ & \ & $\checkmark$ & \ & \ & $\checkmark$ & \ & $\checkmark$ & $\checkmark$ & \ & $\checkmark$ & $\checkmark$ & $\checkmark$ & $\checkmark$ \\ \hline
\end{tabular}
\caption{{All possible combinations of non-zero values for $\nu_k$.} In each case, the symbol $\checkmark$ represents $\nu_k \sim {\cal O}(M_c^{-1})$ while the empty means $\nu_k=0$ at the corresponding fixed point~$z_k$.}
\label{braneplace}
\end{center}
\end{table}
In the following, {we denote} the number of pairs of up- and down-type Higgs KK zero-modes by $n_h \, (=M_\eta)$ and the number of $T^2/Z_2$ fixed points filled with nonvanishing $\mu$-terms ($\nu_k \ne 0$) by $n_\nu \, (\le 4)$. 

It is important to note that we generically encounter the class in which $r<n_h$, where $r$ is the rank of the $\mu$-matrix~(\ref{mumatrix}), where massless Higgs pairs remain in this class. We will find that such a class corresponds to the cases with $n_h>n_\nu$, those are analyzed in the next section~\ref{class2}. In the remaining part of this section, we study the other cases with $n_h \le n_\nu$, where the $\mu$-matrix~(\ref{mumatrix}) has the full rank ($r=n_h$) and all the Higgs pairs become massive due to the localized $\mu$-terms~(\ref{6Dmuterm}). 
{It is not an easy task to diagonalize the $\mu$-matrix~(\ref{mumatrix}) for arbitrary choices of $M$ and $\nu_k$.}
We first show some simple cases for $n_h \le n_\nu$ with $n_h=1,2$ which can be treated analytically in Subsection~\ref{class1analytic}, and then more general cases for $n_h \le n_\nu$ numerically in Subsection~\ref{class1numerical}. In the following, we set ${\rm Re}\,\tau=0$ for simplicity.

\subsection{Analytic calculations for $n_h \le n_\nu$ with $n_h=1,2$}
\label{class1analytic}
First, we study the simplest cases where the single generation of Higgs pair arises ($n_h=1$). The combinations of $M$ and $\eta$ for $H_u$ to realize $n_h=1$ are found in Tab.~\ref{T2Z2generation} as $(M,\eta)=(1,+1)$, $(M,\eta)=(3,-1)$, and $(M,\eta)=(4,-1)$. Although it can be shown that all the fifteen cases shown in Tab.~\ref{braneplace} yield a full rank $\mu$-matrix, we treat the first four cases (Case~1\,--\,4), that is, the localized $\mu$-term is introduced at one of four fixed points ($n_\nu=1$), for simplicity. Eq.~(\ref{mumatrix}) is no longer a matrix and simplified as 
\begin{eqnarray}
\mu_{\rm eff} &=& \nu_k\left(\rho_k^M\right)^2\Theta^{K=0,M}_{T^2}(z_k)\left(\Theta^{L=0,M}_{T^2}(z_k)\right)^*. 
\end{eqnarray}
{Here and hereafter,} we consider a pure imaginary complex structure of $T^2$ as $\tau=i{{\rm Im}\,\tau}$ with the real {and positive} parameter ${{\rm Im}\,\tau}$, and {also} parameterize the position~$z_k$ of the single fixed point~$k$ by $p,q=0,1$ as in Eq.~(\ref{pqdef}) where $\nu_k \ne 0$ is {located}. 
With these parameterizations, the wavefunction~(\ref{wavefuncT2}) on $T^2$ evaluated at $z=z_k$ is denoted by 
\begin{eqnarray}
\Theta^{K=0,M}_{T^2}\left(\frac{p+iq{{\rm Im}\,\tau}}{2}\right) 
&=& \mathcal{N}_Me^{i\frac{\pi Mpq}{4}}\sum_l(-1)^{Mpl}
e^{-\pi M{{\rm Im}\,\tau}\left(l+\frac{q}{2}\right)}, 
\end{eqnarray}
and the effective $\mu$-parameter can be expressed as 
\begin{eqnarray}
\mu_{\rm eff} &=& \nu_k\left(\rho_k^M\right)^2\mathcal{N}_M^2
\left|\sum_l(-1)^{Mpl}e^{-\pi M{{\rm Im}\,\tau}\left(l+\frac{q}{2}\right)^2}\right|^2.
\label{mueff}
\end{eqnarray}
Since the last factor in Eq.~(\ref{mueff}) has a Gaussian-dependence on the index $l$ and a summation over {all integers ($-\infty < l < \infty$)}, $l=0$ give the dominant contribution for $q=0$, while both $l=0$ and $1$ are dominant for $q=1$. {Then, the dominant contribution of $\mu_{\rm eff}$ in Case 1 and 2 ($q = 0$) does not depend on the parameter~${{\rm Im}\,\tau}$, while it depends on the parameter~${{\rm Im}\,\tau}$ in Case 3 and 4 ($q = 1$).} Thus, we estimate the approximate value of $\mu_{\rm eff}$ by the dominant contribution in the following computations. 

For $(M,\eta)=(1,+1)$, the effective $\mu$-parameters~(\ref{mueff}) in Case~1\,--\,4 are estimated as in Tab.~\ref{case1-4}. 
\begin{table}[t]
\begin{center}
\begin{tabular}{|c|c|}\hline
{\rm Case} & $\mu_{\rm eff}/2\nu_k\mathcal{N}_1^2$\\ \hline
$1$ & $1$\\
$2$ & $1$\\
$3$ & $4e^{-\frac{\pi}{2}{{\rm Im}\,\tau}}$\\
$4$ & $0$\\ \hline
\end{tabular}
\caption{Approximate values of the effective $\mu$-parameter~(\ref{mueff}) in Case~1\,--\,4 for $(M,\eta)=(1,+1)$. {The approximate values do not depend on the parameter~${{\rm Im}\,\tau}$ in Case~1 and 2, while they depend on the parameter~${{\rm Im}\,\tau}$ in Case~3 because of the $q$ dependence of the effective $\mu$-parameter~(\ref{mueff}). In Case~4, $\mu_{\rm eff}$ exactly vanishes because the wavefunction at the fixed point $k=4$ vanishes with the odd number of fluxes as shown in Eq.~(\ref{rho}).} We find that a TeV scale $\mu$-parameter can be obtained in Case~3 with ${\rm Im}\,\tau \simeq 20$, even if the compactification scale~$M_c$ is of ${\cal O}(10^{16})$ GeV.}
\label{case1-4}
\end{center}
\end{table}
In Case~1 and 2, the dominant contribution does not depend on the parameter~${{\rm Im}\,\tau}$ {as mentioned above}, while $\mu_{\rm eff}$ exactly vanishes in Case~4 because the wavefunction at the fixed point $k=4$ vanishes with the odd number of fluxes as shown in Eq.~(\ref{rho}). On the other hand, in Case 3, it is remarkable that the ${\rm Im}\,\tau$-dependence of the effective $\mu$-parameter remains in a form of the exponential factor, that can provide a solution to the so-called $\mu$-problem with a suitable value of ${{\rm Im}\,\tau}$. The coefficient $2\nu_k\mathcal{N}_1^2$ has the mass dimension $+1$ and its natural order will be of $O(M_c)$, because the compactification scale~$M_c$ plays a role of a cut-off scale in the effective theories of zero-modes. Then, if we consider that the {compactification} scale is {around} the unification scale~$\sim 10^{16}$ GeV, the effective $\mu$-parameter in Case 3 can be estimated as
\begin{eqnarray}
\mu_{\rm eff} &\sim& 4e^{-\frac{\pi}{2}{{\rm Im}\,\tau}}M_c \ \sim \ 10^{3}\ {\rm GeV}, 
\label{eq:tevmu}
\end{eqnarray}
with ${{\rm Im}\,\tau}=20$. Therefore, we can obtain a TeV scale $\mu$-parameter\footnote{Note that the area~${\cal A}$ of torus is proportional to ${\rm Im}\,\tau$. Hence the wavefunction normalization factor~$\mathcal{N}_M$ in Eq.~(\ref{mueff}) and the compactification scale (in the $y^6$-direction) have mild-dependencies on ${\rm Im}\,\tau$, but their effect can be ignored compared with the one from the exponential factor~$e^{-\frac{\pi}{2}{{\rm Im}\,\tau}}$ shown explicitly in Eq.~(\ref{eq:tevmu}).} if the imaginary part of the complex structure moduli is fixed suitably. {In this case, {only} the fixed point $z_3$ can realize the TeV scale effective {$\mu$-parameter. 
However, which fixed points {can} reproduce a TeV scale $\mu$-parameter} depends on choices of $(M, \eta)$ and the gauge of the flux background.} Similar properties are observed for $(M,\eta)=(3,-1)$ and a TeV scale $\mu$-parameter can be realized in Case~4 with ${{\rm Im}\,\tau} \simeq 60$, while the effective $\mu$-parameter always vanishes in Case~1\,--\,4 for $(M,\eta)=(4,-1)$. 

Next, we describe the case where two generations of Higgs pair arise ($n_h=2$), showing 
the analysis for $(M,\eta)=(2,+1)$ as a simple example. Two generations are labeled by $K=0,1$ and the effective $\mu$-matrix becomes $2\times2$ one, whose components are evaluated by 
\begin{eqnarray}
\Theta^{K=0,M=2}_{T^2}\left(\frac{p+iq{{\rm Im}\,\tau}}{2}\right)
&=& \mathcal{N}_2e^{i\frac{\pi pq}{2}}
\sum_le^{-2\pi{{\rm Im}\,\tau}\left(l+\frac{q}{2}\right)^2}, \\
\Theta^{K=1,M=2}_{T^2}\left(\frac{p+iq{{\rm Im}\,\tau}}{2}\right)
&=& \mathcal{N}_2(-1)^{p}e^{i\frac{\pi pq}{2}}
\sum_le^{-2\pi{{\rm Im}\,\tau}\left(l+\frac{q+1}{2}\right)^2}.
\end{eqnarray}
To obtain a $2 \times 2$ full rank $\mu$-matrix, we restrict ourselves to the last eleven cases (Case~5\,--\,15) in Tab.~\ref{braneplace}. 
For example, in Case 5, the $\mu$-matrix can be estimated, in the same approximation as
the above single-generation cases for Gaussian factors, as
\begin{eqnarray}
\mu_{\rm eff} &\sim& \mathcal{N}_2^2\left(
\begin{array}{cc}
\nu_1+\nu_2 & 2\left(\nu_1-\nu_2\right)e^{-\frac{\pi}{2}{{\rm Im}\,\tau}}\\
2\left(\nu_1-\nu_2\right)e^{-\frac{\pi}{2}{{\rm Im}\,\tau}} & 
4\left(\nu_1+\nu_2\right)e^{-\pi{{\rm Im}\,\tau}}
\end{array}
\right)\nonumber\\
\label{analyticmumatrix}
&=& \mathcal{N}_2^2\left(\nu_1+\nu_2\right)\left(
\begin{array}{cc}
1 & 2Ce^{-\frac{\pi}{2}{{\rm Im}\,\tau}}\\
2Ce^{-\frac{\pi}{2}{{\rm Im}\,\tau}} & 4e^{-\pi{{\rm Im}\,\tau}}
\end{array}
\right), 
\end{eqnarray}
where $C\equiv\left(\nu_1-\nu_2\right)/\left(\nu_1+\nu_2\right)$.  
If the parameter~${{\rm Im}\,\tau}$ satisfy $e^{-\frac{\pi}{2}{{\rm Im}\,\tau}} \ll 1$, the matrix~(\ref{analyticmumatrix}) can be diagonalized as
\begin{eqnarray}
\mu_{\rm eff}^{\rm diag} &\sim& 
\mathcal{N}_2^2\left(\nu_1+\nu_2\right)\left(
\begin{array}{cc}
4\left(1-C^2\right)e^{-\pi{{\rm Im}\,\tau}} & 0\\
0 & 1
\end{array}
\right), 
\label{diagmu}
\end{eqnarray}
where we find hierarchical eigenvalues. 

As in the above argument for single-generation cases, the natural order of the overall factor $\mathcal{N}_2^2\left(\nu_1+\nu_2\right)$ in Eq.~(\ref{diagmu}) will be of $O(M_c)$ and we consider $M_c \sim 10^{16}$ GeV again. Therefore, one of two eigenvalues in Eq.~(\ref{diagmu}) can be of ${\cal O}(10^3)$ GeV with ${{\rm Im}\,\tau} \simeq 10$, while {another} is of ${\cal O}(10^{16})$ GeV. It is a quite interesting feature of this model that {only} a single pair of Higgs {multiplets} can obtain a TeV scale $\mu$-term, while {another} pair completely decouples {with low scale physics and receives} a supersymmetric mass of the order of the compactification scale. We expect that this type of mechanism will be utilized to effectively realize the minimal supersymmetric standard model (MSSM) with a TeV scale $\mu$-term in a certain class of magnetized orbifold models.\footnote{We here comment that, unlike the imaginary part, the real part of the complex structure moduli~${\rm Re}\,\tau$ does not contribute to generate a hierarchical structure, but it can yield CP-violating phases~\cite{Kobayashi:2016qag} instead.} Although we could obtain this result analytically in Case~5, it is in general hard to perform similar analyses for the other cases. Hence, we study them numerically in the next subsection.

\subsection{Numerical analysis for $n_h \le n_\nu$}
\label{class1numerical}
Finally, we analyze the effect of localized $\mu$-terms~(\ref{6Dmuterm}) in detail numerically. In this section, we study the cases with $n_h \le n_\nu$ where full rank $\mu$-matrices are obtained. There are seven patterns of fluxes~{$M=1,2,\cdots,7$},  those yield less than five generations of Higgs pairs in 4D effective theories, and all the possibilities are shown in Tab.~\ref{numericalpattern}. 
\begin{table}[t]
\begin{center}
\begin{tabular}{|c|c|c|c|c|c|c|c|}\hline
$M$ & $1$ & $2$ & $3$ & $4$ & $5$ & $6$ & $7$\\ \hline
{$\#$ of Higgs pair} & $1$ & $2$ & $2$ & $3$ & $3$ & $4$ & $4$\\ \hline
Case in Tab.~\ref{braneplace} & $1$\,--\,$15$ & $5$\,--\,$15$ & $5$\,--\,$15$ & $11$\,--\,$15$ & $11$\,--\,$15$ & $15$ & $15$\\ \hline
\end{tabular}
\caption{All the possibilities yielding less than five generations of Higgs pairs. To have full rank $\mu$ matrices, the number of fixed points filled with $\nu_k \ne 0$ must be equal to or more than the number of Higgs zero-modes, restricting the possible combinations of $\nu_k$ in Tab.~\ref{braneplace}.}
\label{numericalpattern} 
\end{center}
\end{table}
We have to consider that the Higgs fields basically have the even parity~$\eta=+1$, since the wavefunctions of the parity odd modes~$\eta=-1$ mostly vanish at the fixed points {as followed from Eq.~(\ref{nazo})}, and then the full rank $\mu$-matrix does not appear except for the single-generation Higgs cases.

The numerical results of all possibilities are provided in Tabs.~\ref{oneHiggsresult} and \ref{multiHiggsresult} for single- and multi-generation Higgs cases, respectively. 
\begin{table}[t]
\begin{center}
\begin{tabular}{|c|ccc|}\hline
$(M,\eta)$ & $(1,+1)$ & $(3,-1)$ & $(4,-1)$ \\ \hline
Case $1$ & / & -- & --\\
Case $2$ & / & -- & --\\
Case $3$ & $20$ & -- & --\\
Case $4$ & -- & $60$ & --\\
Case $5$ & / & -- & --\\
Case $6$ & / & -- & --\\
Case $7$ & / & $60$ & --\\
Case $8$ & / & -- & --\\
Case $9$ & / & $60$ & --\\
Case $10$ & $20$ & $60$ & --\\
Case $11$ & / & -- & --\\
Case $12$ & / & $60$ & --\\
Case $13$ & / & $60$ & --\\
Case $14$ & / & $60$ & --\\
Case $15$ & / & $60$ & --\\ \hline
\end{tabular}
\caption{The approximate values of ${{\rm Im}\,\tau}$ with which the single pair of Higgs has a TeV scale $\mu$-parameter in the effective theory. {Hyphens {(--)} and slashes {(/)} represent that one or more Higgs pairs remain massless and all the pairs receive heavy masses of $\mathcal{O}(M_c)$, respectively.}
}
\label{oneHiggsresult}
\end{center}
\end{table}
\begin{table}[t]
\begin{center}
\begin{tabular}{|c|cccccc|}\hline
$(M,M_\eta)$ & $(2,2)$ & $(3,2)$ & $(4,3)$ & $(5,3)$ & $(6,4)$ & $(7,4)$\\ \hline
Case $5$ & $10$ & $15$ & -- & -- & -- & --\\
Case $6$ & / & $60$ & -- & -- & -- & --\\
Case $7$ & / & -- & -- & -- & -- & --\\
Case $8$ & / & $60$ & -- & -- & -- & --\\
Case $9$ & / & -- & -- & -- & -- & --\\
Case $10$ & $10$ & -- & -- & -- & -- & --\\
Case $11$ & / & $60$ & $20$ & $25$ & -- & --\\
Case $12$ & / & $15$ & $20$ & -- & -- & --\\
Case $13$ & / & $60$ & $20$ & -- & -- & --\\
Case $14$ & / & $60$ & $20$ & -- & -- & --\\
Case $15$ & / & $60$ & $20$ & $25$ & multi & --\\ \hline
\end{tabular}
\caption{The approximate values of ${{\rm Im}\,\tau}$ with which only a single pair of Higgs has a TeV scale $\mu$-parameter while others receive masses of ${\cal O}(M_c)$ in the effective theory. {Hyphens {(--)}, slashes {(/)}, and ``multi" represent that one or more Higgs pairs remain massless, all the pairs receive heavy masses of $\mathcal{O}(M_c)$, and more than one Higgs pairs have a TeV scale $\mu$-parameter in the effective theory, respectively.}
}
\label{multiHiggsresult}
\end{center}
\end{table}
The numbers in these tables express the approximate values of ${{\rm Im}\,\tau}$ with which only a single pair of Higgs has a TeV scale $\mu$-parameter in the effective theory. On the other hand, {hyphens, slashes, and ``multi" represent that one or more Higgs pairs remain massless, all the pairs receive heavy masses of $\mathcal{O}(M_c)$, and more than one Higgs pairs have a TeV scale $\mu$-parameter in the effective theory, respectively.} We also show the ${\rm Im}\,\tau$-dependence of $\mu_{\rm eff}$ for typical cases in Fig.~\ref{123-genresult}. 
\begin{figure}[t]
\centering
\begin{minipage}{0.48\textwidth}
\centering
\includegraphics[width=\textwidth]{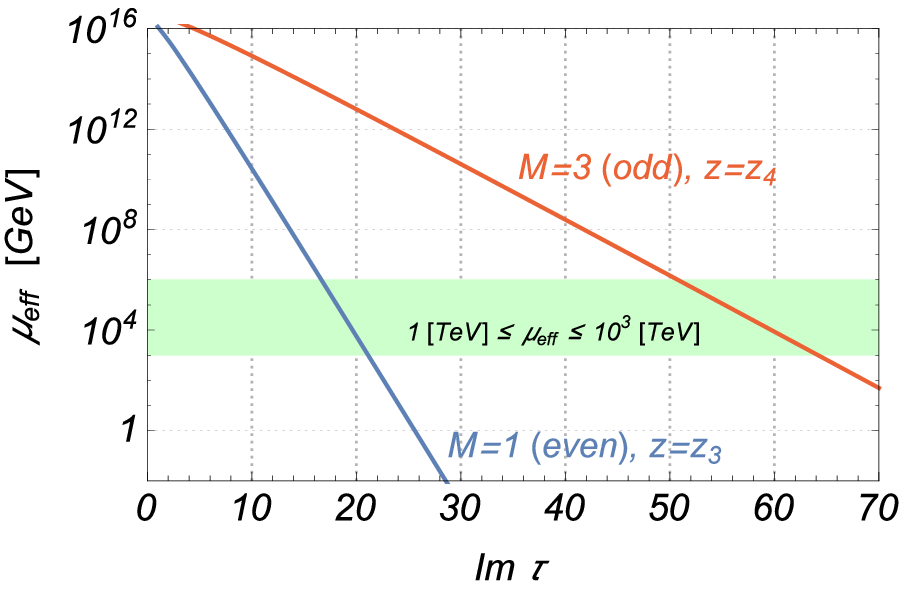}
\\[5pt] {\bf (a)}
\end{minipage}
\hfill 
\begin{minipage}{0.48\textwidth}
\centering
\includegraphics[width=\textwidth]{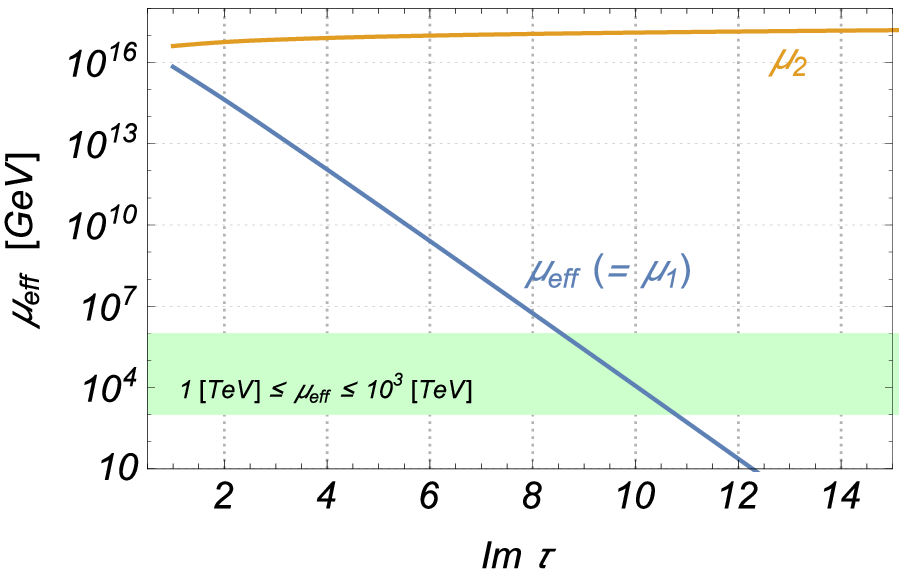}
\\[5pt] {\bf (b)}
\end{minipage}
\\[10pt]
\begin{minipage}{0.48\textwidth}
\centering
\includegraphics[width=\textwidth]{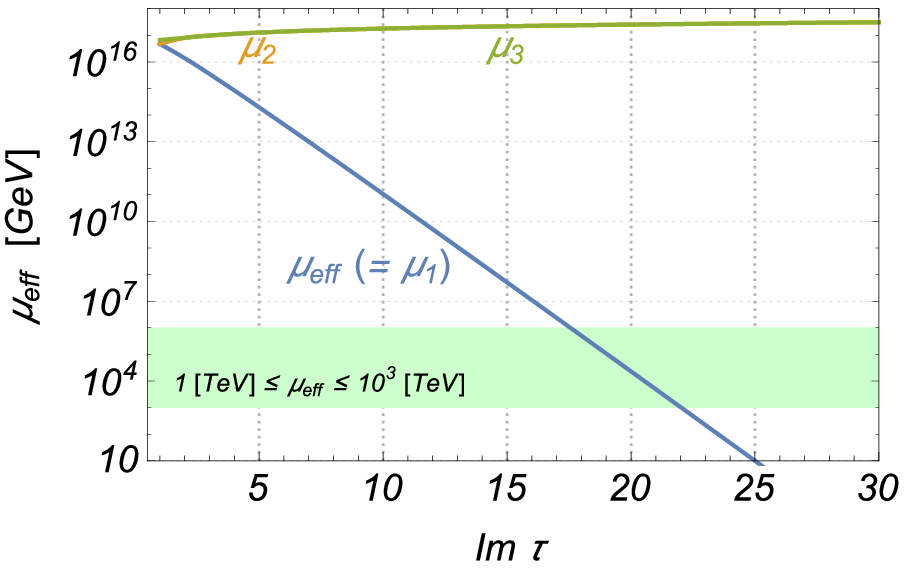}
\\[5pt] {\bf (c)}
\end{minipage}
\caption{The ${\rm Im}\,\tau$-dependences of the lowest three eigenvalues~$\mu_i$ ($i=1,2,3$) of the $\mu$-matrix~(\ref{mumatrix}). We set $\nu_k^{-1}=M_c=10^{16}$ GeV for $\nu_k \ne 0$ and the other parameters are chosen as {\bf (a)}~$(M,\eta)=(1,+1)$ in Case 3 (blue line) and $(M,\eta)=(3,-1)$ in Case 4 (red line), {\bf (b)}~$(M,\eta)=(2,+1)$ in Case 5, and {\bf (c)}~$(M,\eta)=(4,+1)$ in Case 15. The figure~{\bf (a)} corresponds to the single Higgs cases where $\mu_2$ and $\mu_3$ are absent, while the figures~{\bf (b)} and {\bf (c)} correspond to the multi-Higgs cases. In the light green region, eigenvalues reside in between $1$~TeV and $10^3$~TeV.}
\label{123-genresult}
\end{figure}
The upper two figures are the numerical results from single- and two-generation Higgs cases calculated analytically in the previous subsection. The lower figure is the result from one of the three-generation Higgs cases, which has a remarkable structure that, {even though} all the four fixed points are filled with $\nu_k \ne 0$, there {exists only} one TeV scale $\mu$-parameter while the others are of ${\cal O}(M_c)$. 
Therefore, we expect that {the} MSSM with a TeV scale $\mu$-term can be effectively realized in magnetized orbifold models with three-generation Higgs pairs.

Before closing this section, we make some comments on the Yukawa couplings {to the SM quarks and leptons}. The gauge invariance requires $M_1+M_2=M_3$ in the expression of the Yukawa coupling~(\ref{T2yukawa}), where $M_1$ and $M_2$ can be identified with the fluxes felt by left- and right-handed quarks or leptons, respectively, and then $M_3$ determines the number of Higgs generations. To obtain three generations of quarks and leptons with nonvanishing Yukawa couplings based on Tab.~\ref{T2Z2generation}, we find that $M_3=8,9,10,14,15$ or $16$ are allowed by the condition~$M_1+M_2=M_3$~\cite{Abe:2008sx} {unless we suppose nontrivial Wilson line twists}. Thus, the number of Higgs generation cannot be less than five in this case.\footnote{{In an extension to supergravity, it is investigated that $U(1)$ associated with flux background can be broken by St\"uckelberg-like mechanism \cite{Buchmuller:2015eya}. Then, more various possibilities for model constructions would be possible. In the whole of this paper, we implicitly suppose $U(1)$ preserving situations.}} In other words, {since the maximum rank of the $\mu$-matrices {is} four because of the four  fixed points on $T^2/Z_2$,} any model with $n_h \le n_\nu$ ($\le 4$ on $T^2/Z_2$) analyzed in this section cannot have {bulk} Yukawa couplings~(\ref{T2yukawa}). There will be, however, some possibilities to produce them by considering other effects such as higher-order couplings, nonperturbative effects as well as some localized generations and/or Yukawa couplings at the fixed points. Therefore, it is nevertheless important to study multi-Higgs models with less than five generations as has been done in this section. 

On the other hand, the same argument tells us that the cases with $n_h>n_\nu$ can produce nonvanishing Yukawa couplings~(\ref{T2yukawa}), those will be studied in the next section. We will find that the effective $\mu$-matrix cannot be full rank and massless Higgs pairs remain in these cases, those could obtain masses due to, e.g., nonperturbative and/or supersymmetry breaking effects. Therefore, in the next section, we focus on the structure of Yukawa matrices by identifying one of the massless pairs as a Higgs pair {in} the MSSM.

\section{Analysis for $n_h>n_\nu$ and Yukawa matrices}
\label{class2}
In the previous section, we investigated the cases with $n_h \le n_\nu$ where the $\mu$-matrices become full rank. In this section, we study the other cases with $n_h>n_\nu$. The number~$n_0$ of zero eigenvalues of the $\mu$-matrix is, in general, determined by the size~$n_h \times n_h$ of the matrix and the number~$n_\nu$ of fixed points filled with {nonvanishing} $\mu$-terms. For instance, the numbers of~$n_0$ for $n_h>n_\nu=4,3$ on $T^2/Z_2$ are shown in Tab.~\ref{zeronumber}, where we find that the effective $\mu$-matrix cannot be full rank, $n_0 \ne 0$, and the number~$n_0$ is determined as $n_0=n_h-n_\nu$. 
\begin{table}[t]
\begin{center}
\begin{tabular}{|c|ccccccc|}\hline
$M$ & $8$ & $9$ & $10$ & $11$ & $12$ & $13$ & $\cdots$\\ \hline
$n_h=M_+$ & $5$ & $5$ & $6$ & $6$ & $7$ & $7$ & $\cdots$\\
$n_\nu$ & $4$ & $3$ & $4$ & $3$ & $4$ & $3$ & $\cdots$\\ \hline
$n_0$ & $1$ & $2$ & $2$ & $3$ & $3$ & $4$ & $\cdots$\\ 
\hline
\end{tabular}
\caption{The number~$n_0$ of zero eigenvalues of the $\mu$-matrix for $n_h>n_\nu=4,3$ on $T^2/Z_2$, where the Higgs fields are assigned as parity even~($\eta=+1$) and then $n_h=M_+$.}
\label{zeronumber}
\end{center}
\end{table} 
This result tells us that the original number~$n_h$ of massless Higgs pairs without the localized sources is effectively reduced by the number~$n_\nu$ of localized $\mu$-terms, that may help to reduce the number of multiple (in general too many) Higgs zero-modes in magnetized orbifold models. Note that $n_\nu=n_h-n_0$~Higgs fields receive heavy masses of {${\cal O}(M_c)$} in this case, then decouple from $n_0$~zero-modes at low energies. 

If we consider that one of the remaining massless pairs plays the roles of MSSM Higgs (super)fields $H_u$ and $H_d$, it is worth analyzing the Yukawa couplings with quarks and leptons derived as overlap integrals of their wavefunctions, those are also determined by magnetic fluxes. In previous works~\cite{Abe:2012fj,Abe:2014vza,Abe:2016jsb}, it was shown that semi-realistic patterns of mass ratios and mixing angles among three generations of quarks and leptons are obtained on magnetized orbifolds, by {\it assuming} a certain linear combination of multiple Higgs zero-modes to be the MSSM one and {the} others are somehow heavy. Then, it would be interesting if the particular linear combination is {\it selected} as a massless direction by localized {$\mu$-terms} at the fixed points that realizes a realistic pattern of quark and/or lepton masses and mixings. {In this case, as mentioned at the end of the previous section, some effects such as nonperturbative ones {in addition to} the localized sources are required to obtain the $\mu$-term for the MSSM Higgs pairs (and those for the other massless pairs if $n_0>1$), which we implicitly assume in the following arguments.}\footnote{Changing the orbifold structure and/or higher-dimensional embeddings may provide another way to avoid massless Higgses, though it is quite nontrivial to consider that they don't affect the essential part of our mechanism realized on $T^2/Z_2$. We remark that the complex structure is fixed on $T^2/Z_N$ for $N>2$, and it is hard to derive a hierarchical structure in this case within the framework of $T^2$.}


Now we study Yukawa couplings in the case that $(H_u,H_d)$ appears as the single massless pair among all the eigenstates of $\mu$-matrix, i.e. $n_0=1$, with three generations of left- and {right}-handed fermions (quarks or leptons) charged under the fluxed $U(1)$. From Tabs.~\ref{T2Z2generation} and \ref{zeronumber}, we find that the unique combination of fluxes $(M_1,M_2,M_3)=(4,4,8)$ satisfies these conditions and at the same time allows nonvanishing Yukawa coupling constants~(\ref{T2yukawa}), where $M_1$, $M_2$ and $M_3$
are felt by left-, right-handed fermion and Higgs fields, respectively. 
In this case, the $\mu$-matrix~(\ref{mumatrix}) is determined by $M=8$ wavefunction evaluated at the fixed points, 
\begin{eqnarray}
\Theta^{K,M=8}_{T^2}(z_k)
&=&(-1)^{pK}\Theta^{K+4q,M=8}_{T^2}(z_1) \ = \ (-1)^{pK}u^{K+4q}, 
\label{defu}
\end{eqnarray}
where $K=0,1,2,3,4$ for $M_\eta=M_+=5$ in Tab.~\ref{T2Z2generation} and 
the fixed points $z_k$ labeled by $k=1,2,3,4$ correspond to $(p,q)=(0,0)$, $(0,1)$, $(1,0)$, $(1,1)$, respectively, as in Eq.~(\ref{pqdef}). 
Here we define $u^K \equiv \Theta^{K,M=8}_{T^2}(z_1)$ 
which has a periodicity $u^{K+8}=u^K$. We find 
\begin{eqnarray}
\mu_{\rm eff}^{KL} 
&=& \sum_{k=1}^42\nu_k\Theta^{K,M=8}_{T^2}(z_k)\left(\Theta^{L,M=8}_{T^2}(z_k)\right)^* 
\ = \ \sum_{k=1}^42\nu_k(-1)^{pK}u^{K+4q}\left((-1)^{pL}u^{L+4q}\right). 
\nonumber
\end{eqnarray}
The eigenvector~$v^K$ with the zero eigenvalue which satisfies 
$\displaystyle \sum_{L=0}^4 \mu_{\rm eff}^{KL} v^L=0$ is determined by 
\begin{eqnarray}
\sum_{K=0}^4(-1)^{pK}u^{K+4q}v^K &=& 0 \qquad ^\forall (p,q), 
\label{eigeneq}
\end{eqnarray}
which can be expressed in a matrix form as
\begin{eqnarray}
\left(
\begin{array}{ccccc}
u^0 & u^1 & u^2 & u^3 & u^4\\
u^0 & -u^1 & u^2 & -u^3 & u^4\\
u^4 & u^3 & u^2 & u^1 & u^0\\
u^4 & -u^3 & u^2 & -u^1 & u^0
\end{array}
\right)\left(
\begin{array}{c}
v^0\\
v^1\\
v^2\\
v^3\\
v^4
\end{array}
\right) \ = \ 0.
\end{eqnarray}
The normalized solution of the above equation is
\begin{eqnarray}
v^K &=& \frac{1}{\sqrt{\omega^2+2}}\left(1,\ 0,\ \omega,\ 0,\ 1\right), 
\nonumber
\end{eqnarray}
where 
\begin{eqnarray}
\omega &\equiv& -\frac{u^0+u^4}{u^2}, 
\nonumber
\end{eqnarray}
which depends on the imaginary part of the complex structure moduli, 
${{\rm Im}\,\tau}$. 

The Yukawa matrix for three generation fermions is
\begin{eqnarray}
Y^{IJ} &=& \sum_{K=0}^4Y^{IJK}v^K 
\ = \ \frac{1}{\sqrt{\omega^2+2}}\left(
\begin{array}{ccc}
y_a+y_e & 0 & \omega y_c\\
0 & \frac{\omega}{\sqrt{2}}\left(y_a+y_e\right)+2y_c & 0\\
\omega y_c & 0 & y_a+y_e
\end{array}
\right), 
\label{yukawa_matrix}
\end{eqnarray}
where we adopt the notations of Ref.~\cite{Abe:2008sx}, 
\begin{eqnarray}
\begin{array}{ll}
y_a \ = \ \eta_0+2\eta_{32}+\eta_{64},\ & 
y_b \ = \ \eta_4+\eta_{28}+\eta_{36}+\eta_{60}, \\
y_c \ = \ \eta_8+\eta_{24}+\eta_{40}+\eta_{56}, \ & 
y_d \ = \ \eta_{12}+\eta_{20}+\eta_{44}+\eta_{52}, \\
y_e \ = \ 2\eta_{16}+2\eta_{48}, 
\end{array}
\end{eqnarray}
and 
\begin{eqnarray}
\eta_N=\vartheta
\begin{bmatrix}
N/128\\
0  
\end{bmatrix}
\left(0,128\tau\right).
\end{eqnarray}
We can approximate the above Yukawa matrix as
\begin{eqnarray}
Y^{IJ} &\sim& \frac{1}{\sqrt{\omega^2+2}}\left(
\begin{array}{ccc}
\eta_0 & 0 & \omega\eta_8\\
0 & \frac{\omega}{\sqrt{2}}\eta_0 & 0\\
\omega\eta_8 & 0 & \eta_0
\end{array}
\right),
\end{eqnarray}
and the eigenvalues are described as
\begin{eqnarray}
\left|\eta_0\pm\omega\eta_8\right|^2, \qquad 
\frac{1}{2}\left|\omega\right|^2\left|\eta_0\right|^2. 
\label{yukawa_ev}
\end{eqnarray}

Note that these eigenvalues depend on {only} the imaginary part of the complex structure {modulus}~$\tau$ {since we assume the real part of the complex structure is zero for simplicity}. The ${\rm Im}\,\tau$-dependences of fermion mass ratios generated by the Yukawa matrix~(\ref{yukawa_matrix}) are calculated numerically and shown in Fig.~\ref{hierarchygraph}. {In this figure, $m_i$ represents the mass of {$i$-th} generation fermion ($m_1<m_2<m_3$) and the blue and yellow lines correspond to the ratios $m_1/m_3$ and $m_2/m_3$, respectively. The observed values for up- and down-type quarks are shown by horizontal {dashed and dash-dotted} lines where $m_u$, $m_c$, $m_t$, $m_d$, $m_s$ and $m_b$ denote up, charm, top, down, strange and bottom quark masses, respectively.}
\begin{figure}[t]
\centering
\includegraphics[width=0.5\textwidth]{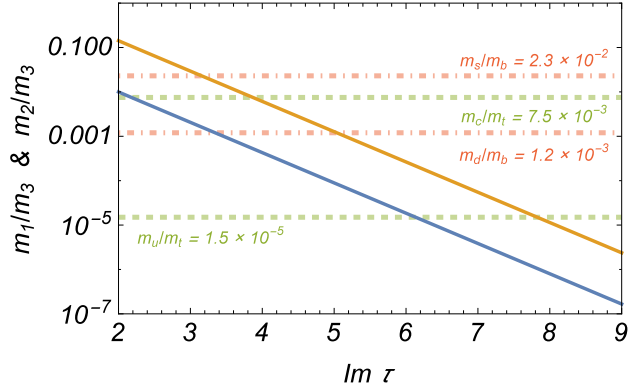}
\caption{The ${\rm Im}\,\tau$-dependences of fermion mass ratios generated by the Yukawa matrix~(\ref{yukawa_matrix}), where $m_i$ represents the mass of {$i$-th} generation fermion ($m_1<m_2<m_3$). The blue and yellow lines correspond to the ratios $m_1/m_3$ and $m_2/m_3$, respectively. The observed values for up- and down-type quarks are shown by horizontal {dashed and dash-dotted} lines. We find ${{\rm Im}\,\tau} \simeq 3.2$ produces the observed mass ratios of down-type quarks. {Note that we set ${\rm Re}\,\tau=0$ for simplicity.}}
\label{hierarchygraph}
\end{figure}
For ${{\rm Im}\,\tau}=3.2$, the three eigenvalues~(\ref{yukawa_ev}) are estimated as
\begin{eqnarray}
\left(1.5\times10^{-3},\ 2.1\times10^{-2},\ 1\right), 
\nonumber
\end{eqnarray}
those are compared with the observed mass ratios, 
\begin{eqnarray}
\left(m_d,m_s,m_b\right)/m_b 
&\sim& \left(1.2\times10^{-3},\ 2.3\times10^{-2},\ 1\right), 
\nonumber
\end{eqnarray}
Therefore, the mass hierarchy of down-type quarks can be produced, if the complex structure {modulus} of $T^2/Z_2$ is fixed as ${{\rm Im}\,\tau}=3.2$. 
{In this case, the mass hierarchies of up-type quarks and charged leptons cannot be produced in any {value of} ${{\rm Im}\,\tau}$, but if we consider other effects as mentioned in Sec.~\ref{class1} there will be some possibilities to produce them.}

\section{Effects of KK excitation modes}
\label{sec:KKeffects}
In this section, we show that the effects of KK excitation modes on the lowest eigenstates of $\mu$-matrix are negligible based on numerical analyses. In our 6D models, $H_u$ and $H_d$ {originate from} hypermultiplets $(H_u,H_u^C)$ and $(H_d,H_d^C)$, respectively. Note that $H_u^C$ and $H_d^C$ are conjugate representations of $H_u$ and $H_d$, respectively, under all gauge groups including the fluxed $U(1)$ 
{and have opposite sign of $Z_2$-parity to $H_u$ and $H_d$, respectively}. 
Therefore, when $H_u$ and $H_d$ feel nonvanishing magnetic fluxes yielding zero-modes, the wavefunctions of $H_u^C$ and $H_d^C$ zero-modes are not normalizable and then eliminated. 

As mentioned in the previous sections, we implicitly assume that our 6D models are embedded into 10D that normally implies the mass terms of hypermultiplets are absent in the bulk. In this case with $M>0$ and $\nu_k \ne 0$ \ $^\exists k$ in Eq.~(\ref{6Dmuterm}), the mass terms in the 4D effective theory are written in ${\cal N}=1$ superspace as
\begin{eqnarray}
\mathcal{L}^{\rm eff}_{\rm mass} 
&=& \int d^2\theta\left[\left(H_u^{(m,K)},H_d^{C(m',K')}\right)\left(
\begin{array}{cc}
\mu_{\rm eff,{\eta=+1}}^{(m,K),(n,L)} &m_n\delta^{(m,K),(n',L')}\\
m_n\delta^{(m',K'),(n,L)} &{\mu_{\rm eff,\eta=-1}^{(m',K'),(n',L')}}
\end{array}
\right)\left(
\begin{array}{c}
H_d^{(n,L)}\\
H_u^{C(n',L')}
\end{array}
\right)+{\rm h.c.}\right], 
\nonumber \\
\label{effmass}
\end{eqnarray}
with 
\begin{eqnarray}
m_n^2 &\equiv& \frac{4\pi M}{\mathcal{A}}n, \qquad 
\mu_{\rm eff,{\eta}}^{(m,K),(n,L)} 
\ \equiv \ \sum_{k=1}^4\nu_k\Theta_{T^2/Z_2,{\eta}}^{(m,K),M}(z_k)
\left({\Theta}_{T^2/Z_2,{\eta}}^{(n,L),M}(z_k)\right)^*, 
\nonumber
\end{eqnarray}
where $m,n$ ($m',n'$) and $K,L$ ($K',L'$) label the KK modes of $H_{u,d}$ ($H_{d,u}^C$) and their degeneracy, respectively, 
{$\eta$ denotes the $Z_2$-parity of the wavefunctions,} 
and the summation symbols for these indices are suppressed. Note that ${m'},n' \ne 0$ due to the absence of the zero-modes mentioned above. 

In Eq.~(\ref{effmass}), the effective mass matrix including KK excitations is described  as
\begin{eqnarray}
\left(
\begin{array}{cc}
\mu_{\rm eff,{\eta=+1}}^{(m,K),(n,L)} &m_n\delta^{(m,K),(n',L')}\\
m_n\delta^{(m',K'),(n,L)} &{\mu_{\rm eff,\eta=-1}^{(m',K'),(n',L')}}
\end{array}
\right), 
\label{KKmatrix}
\end{eqnarray}
where $\mu_{\rm eff}^{(m=0,K),(n=0,L)}=\mu_{\rm eff}^{KL}$ corresponds to the $\mu$-matrix for zero-modes shown in Eq.~(\ref{mumatrix}). Notice that the full matrix~(\ref{KKmatrix}) has the same feature as the zero-mode one~(\ref{mumatrix}), that is, they are not full rank matrices for $n_h>n_\nu$. Therefore, the massless eigenstates appear in this case. We analyze the effects of KK excitations on the lowest eigenstates (especially the eigenvalues) of the mass matrix~(\ref{KKmatrix}) based on similar analyses performed in Ref.~\cite{Ishida:2017avx,Sakamura:2016kqv}. By {identifying a cut-off scale as} $\Lambda \simeq m_{n=N_{\rm max}}$, the matrix~(\ref{KKmatrix}) is regularized to be a finite one within $n,m,n',m' \le N_{\rm max}$. By diagonalizing the finite matrix numerically, we can derive the $N_{\rm max}$-dependences of the eigenvalues, those are shown in Fig.~\ref{NMaxmueff} for typical parameter choices. 
\begin{figure}[t]
\centering
\begin{minipage}{0.48\textwidth}
\centering
\includegraphics[width=\textwidth]{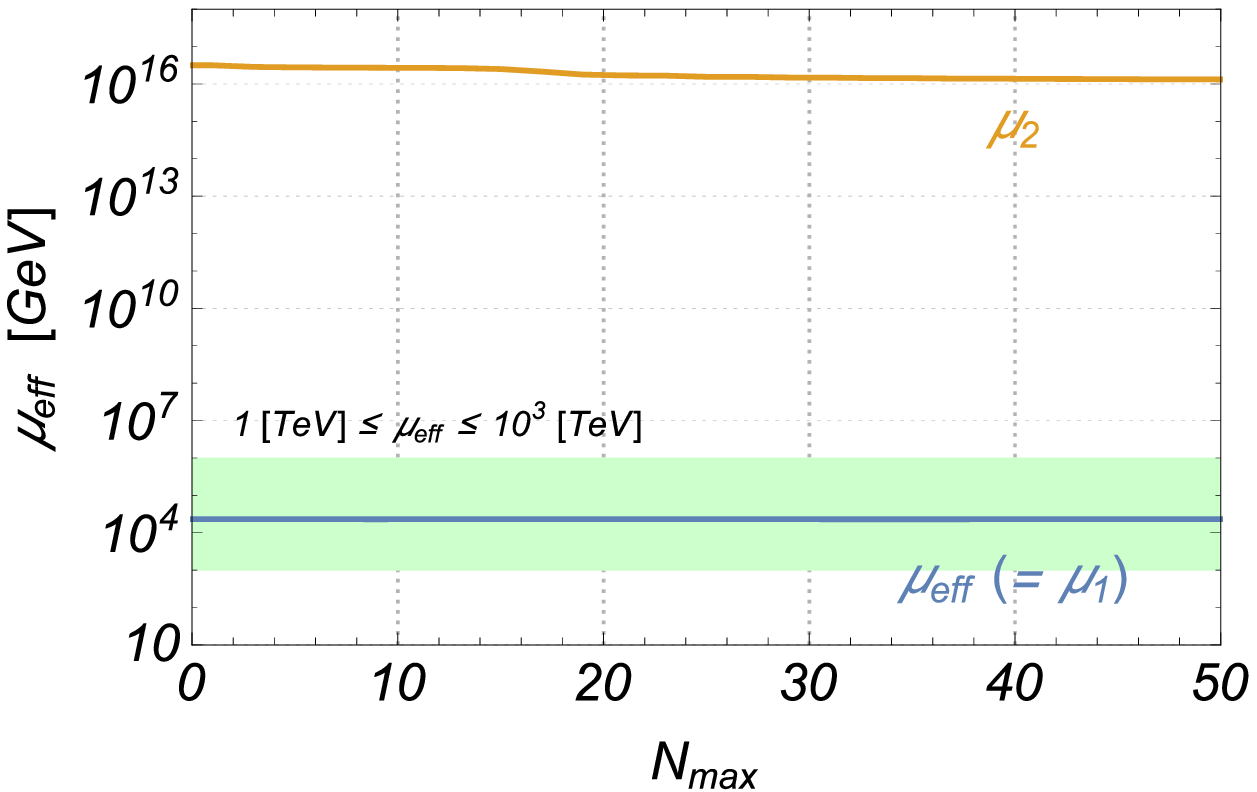}
\\[5pt] {\bf (a)}
\end{minipage}
\hfill
\begin{minipage}{0.48\textwidth}
\centering
\includegraphics[width=\textwidth]{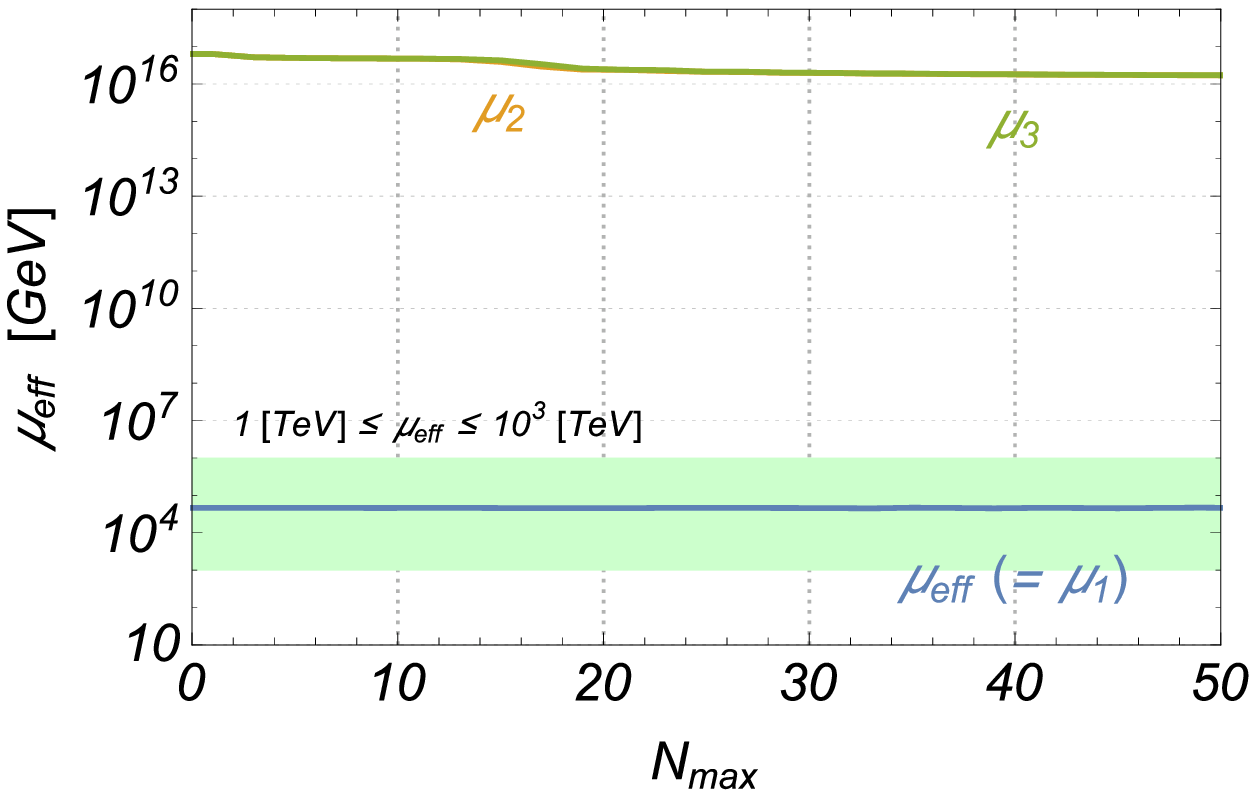}
\\[5pt] {\bf (b)}
\end{minipage}
\caption{The $N_{\rm max}$-dependences of the lowest three eigenvalues~$\mu_i$ ($i=1,2,3$) of the regularized mass matrix~(\ref{KKmatrix}) within $n,m,n',m' \le N_{\rm max}$. We set $M_c=10^{16}$ GeV and the other parameters are chosen as {\bf (a)}~$(M,\eta)=(2,+1)$ in Case 5 with ${{\rm Im}\,\tau}=10$ and $(\nu_1,\nu_2)=(1,-0.5)M_c^{-1}$, and {\bf (b)}~$(M,\eta)=(4,+1)$ in Case 15 with ${{\rm Im}\,\tau}=20$ and $(\nu_1,\nu_2,\nu_3,\nu_4)=(1,-0.5,-0.5,1)M_c^{-1}$. In the light green region, the eigenvalues reside in between $1$~TeV and $10^3$~TeV. {Note that we set ${\rm Re}\,\tau=0$ for simplicity.}}
\label{region}
\label{NMaxmueff}
\end{figure}
From this figure, we confirm that the effects of KK excitations on the lowest eigenvalue are negligible independently {on} the cut-off scale $\Lambda$, which justifies the validity of arguments in the previous sections.

\section{Conclusion}
\label{conclusion}
We have shown in {this paper} how the $\mu$-terms localized at the fixed points affect the degenerated KK zero-modes of Higgs pairs~$(H_u,H_d)$ in magnetized orbifold models on $T^2/Z_2$, where the degeneracy~$n_h$ is determined by the magnetic fluxes~$M$ they feel as well as their orbifold parity~$\eta$. By diagonalizing the $\mu$-matrix~$\mu_{\rm eff}^{KL}$ in the 4D effective theory, we find that the rank of the mass matrix for Higgs zero-modes is raised by the number of fixed points~$n_\nu \le 4$ where nonvanishing $\mu$-terms~$\nu_k \ne 0$ are localized. Thus, $n_0={\rm max}\,(n_h-n_\nu,0)$ pairs remain massless as a result. 

For $n_h \le n_\nu$, all the Higgs zero-modes become massive due to the {localized} $\mu$-terms. Then, we find one of the eigenvalues of $\mu_{\rm eff}^{KL}$ tends to be hierarchically small compared with the others if the imaginary part of the complex structure of $T^2$ is mildly large, due to a Gaussian structure of wavefunctions. The result suggests that the MSSM-like models with a single pair of Higgs possessing a small $\mu$-parameter (compared with the cut-off scale) effectively arise at low energies, even though multiple Higgs zero-modes intrinsically appear as the KK eigenstate with magnetic fluxes. Although nonvanishing Yukawa couplings to three generations of {quarks} and leptons are not allowed in this case with $n_h \le n_\nu$ ($\le 4$ on $T^2/Z_2$), some other perturbative/nonperturbative effects in the bulk and/or at the fixed points are expected to yield them. Therefore, the case mentioned here is interesting, providing a solution to the so-called $\mu$-problem in the derived MSSM-like models with a single Higgs pair. 

For $n_h>n_\nu$, on the other hand, $n_0 \ge 1$ Higgs pairs remain massless. Even in this case, the localized $\mu$-terms play important roles in the phenomenological model building, at any rate, reducing the number of massless pairs by $n_\nu$ compared with the original number of degenerated Higgs KK zero-modes~$n_h$, that is too large in most models. It is remarkable that nonvanishing Yukawa couplings are allowed in this case. Especially, we have found that a single pair of $(H_u,H_d)$, whose VEVs naturally realize observed mass ratios for down-type quarks with a certain value of ${\rm Im}\,\tau$, remain massless among multiple Higgs zero-modes for $n_0=n_h-n_\nu=1$ due to the localized $\mu$-terms. Therefore, at a low energy, we can realize a MSSM-like model, where three generations of quarks and leptons have hierarchical Yukawa couplings to a single pair of $(H_u,H_d)$, generating the observed mass ratios at least for down-type quarks. In some extensions of this case, we would expect an interplay between such localized $\mu$-terms and some other perturbative/nonperturbative effects which generate a $\mu$-term~\cite{Ibanez:2006da,Blumenhagen:2006xt,Abe:2015uma,Kobayashi:2015siy} (and soft supersymmetry breaking terms) for the {remaining} massless pair to develop VEVs, effectively yielding hierarchical quark masses and mixing angles consistent with the {observations}. 

We conclude that, in both cases with $n_h \le n_\nu$ and $n_h>n_\nu$, the localized $\mu$-terms can play important roles in particle physics models constructed on magnetized orbifolds, although these two cases have different phenomenological prospects from each other. It is interesting to consider concrete embeddings of our 6D models into higher-dimensional ones which is just assumed in this paper. From a theoretical point of view, it will be important to study the resolutions of singularities associated with the orbifold fixed points~{\cite{Lust:2006zh,Nibbelink:2007rd,Nibbelink:2009sp,Blaszczyk:2010db}}, that may reveal the origin of localized $\mu$-terms (those are not renormalized perturbatively with ${\cal N}=1$ supersymmetry) where some blow-up modes could be involved. We will study these issues elsewhere.

\section*{Acknowledgement}
The authors would like to thank Yutaka Sakamura for useful discussions.
{Y.T. would like to thank Wilfried Buchm\"uller for instructive comments on this manuscript.} 
H.~A. was supported in part by JSPS KAKENHI Grant Number JP16K05330 and also supported by Institute for Advanced Theoretical and Experimental Physics, Waseda University.
{Y.T. is supported in part by Grants-in-Aid for JSPS Research Fellow (No.~16J04612) and JSPS Overseas Research Fellow (No.~18J60383) from the Ministry of Education, Culture, Sports, Science and Technology in Japan.
}

\bibliographystyle{utphys}
\bibliography{references}

\end{document}